\documentclass[article,twocolumn,showpacs,aps,prb,floatfix]{revtex4-1}
\usepackage{graphicx,subfigure,epsfig,verbatim,psfrag,amsmath,amssymb,color}
\usepackage{indentfirst}
\usepackage{textcomp}
\usepackage{latexsym}
\usepackage{amssymb}
\usepackage{amsmath}
\usepackage{lipsum}
\usepackage{soul}
\usepackage{xcolor}
\makeatletter

\begin{document}
\title{Magnetic properties of 2D topological insulators}
\author{Zewei Chen}
\author{Tai Kai Ng}
\affiliation{Department of Physics, Hong Kong University of Science and Technology, Clear Water Bay, Hong Kong, China}
\begin{abstract}
The effects of Hubbard-type on-site interactions on the BHZ model is studied in this paper  for model parameters appropriate for the HgTe/CdTe quantum well. Within a simple mean field theory we search for plausible magnetic instabilities in the model and find that the ground state becomes {\em ferromagnetic} when the interaction strength between electrons in hole orbital is strong enough. The result can be understood by an approximate mapping of the Hubbard-BHZ model to the one band Hubbard model. The same mapping suggests that the magnetic and/or other ordered phases are more likely to occur in large gap topological insulators whose occupations are close to 1/2 for both electron and hole orbital.
\end{abstract}
\maketitle
\section{Introduction}
Topological phases entered condensed matter physics as early as the Kosterlitz-Thouless (KT) transition, Quantum Spin Chains (QSC) and Integer Quantum Hall effect (IQHE)\cite{KT,HALDANE,IQHE,topo_iqhe} and becomes a heated topic in physics with the discovery of topological insulators (TIs)\cite{RevModPhys.83.1057,Bernevig15122006} , topological superconductors (TSs)\cite{1063-7869-44-10S-S29,PhysRevLett.100.096407,PhysRevB.86.184516} and quantum anomalous hall effect (QAHE) \cite{Chang167}. The role of topology in the physics of non-interacting systems is now basically understood with the completion of topological classification\cite{Schnyder,Kitaev,Shinsei}. However, the role of topology in interacting systems are much more non-trivial and remains to be explored. In 1D system, the classification of TIs and TSs is shown to be changed by interaction\cite{Fidkowski,fidkowski2010effects}. For example, it is well known that the topology of non-interacting TIs cannot be changed without closing the bulk gap. However, this is not true in interacting systems where a first order topological phase transition may occur without closing the bulk gap\cite{Amaricci}. New topological phases protected by interaction are also predicted to exist at dimensions $d>1$\cite{Santos,PhysRevB.82.075125,PhysRevB.82.115125,wang2014classification}.

In this paper, we study the effects of interaction in 2D TIs. We note that there have been many efforts trying to studied TIs with both short range or Coulomb interaction. The electron and hole bands in InAs/GaSb Quantum Well which occupy separate electron and hole layers form a good platform to search for topological exciton formed by Coulomb interaction\cite{PhysRevLett.103.066402,exciton_inas,PhysRevLett.112.146405}. The effects of Hubbard-type interaction in the Bernevig, Hughes and Zhang (BHZ) model\cite{Bernevig15122006} (Hubbard-BHZ model) has also been studied by Miyakoshi {\em et al.}\cite{PhysRevB.87.195133} where a topological phase with anti-ferromagnetic order is predicted. However, the parameters used in their study do not describe the corresponding TI material- HgTe/CdTe quantum well\cite{zhang,1002.2904,type2qsh}. In this paper, we study the Hubbard-BHZ model with parameters appropriate for the HgTe/CdTe quantum well and search for plausible magnetic phases. We find a first order phase transition where the insulating TI phase becomes a ferromagnetic metal phase when the hole-hole interaction is strong enough. 
By mapping approximately the Hubbard-BHZ model to the one band Hubbard model, we explain the origin of the ferromagnetic phase and predict that more exotic phases will appear when the electron occupation is close to 1/2 for both electron and hole orbital.

\section{Model and method}
We consider the BHZ model with Hubbard-type interaction on a square lattice with two orbital $\{|E\sigma\rangle,|H\sigma \rangle \}$ per site. The system is described by a Hamiltonian $H_T=H_{BHZ}+H_{U}$, where $H_{BHZ}=T+H_0$ is the BHZ model with
\begin{subequations}
\begin{align}
H_{0}&=\sum_{\langle i,j\rangle,\sigma} \varepsilon_E C_{i,E,\sigma}^{\dag}  C_{i,E,\sigma} +\varepsilon_H C_{i,H,\sigma}^{\dag}C_{i,H,\sigma}
\end{align}
being the on-site energy term with $\varepsilon_{\tau}$ being the on-site energy for $\tau$ orbital and
\begin{align}
T&=\sum_{\langle i,j \rangle,\sigma}t_E C_{i,E,\sigma}^{\dag}C_{j,E,\sigma}  + t_H C_{i,H,\sigma}^{\dag}C_{j,H,\sigma} \\
&+\sigma t_{EH}[i( C_{i,E,\sigma}^{\dag}C_{i+\hat{x},H,\sigma} -C_{i,E,\sigma}^{\dag}C_{i-\hat{x},H,\sigma}) \\
&+( C_{i,E,\sigma}^{\dag}C_{i+\hat{y},H,\sigma} -C_{i,E,\sigma}^{\dag}C_{i-\hat{y},H,\sigma})]+c.c
\end{align}
\end{subequations}
 describes electron hopping between different lattice sites where $t_{\tau}, t_{EH}$ denotes intra-orbital and inter-orbital hopping, respectively. $C^+(C)_{i,\tau,\sigma}$ creates/annihilates a $\tau$-orbital ($\tau$=E,H) electron with spin $\sigma=\uparrow,\downarrow$ (or $\pm1$) on site $i$. $<i,j>$ denotes nearest neighbor (NN) sites and c.c denotes the hermitian conjugate.
\begin{align}
H_{U}=\sum_{i} \sum_{\tau=E,H} U_{\tau} n_{i,\tau,\uparrow} n_{i,\tau,\downarrow}+\sum_{i} \sum_{\sigma,\sigma'} U_{EH} n_{i,E,\sigma} n_{i,H,\sigma'}
\end{align}
where $U_{\tau} (\tau=E,H), U_{EH}>0$ describe repulsive interaction between electrons and $n_{i,\tau,\sigma}=C_{i,\tau,\sigma}^{\dag} C_{i,\tau,\sigma}$. Fourier transforming, we obtain
\begin{align}
\begin{split}
&H_{BHZ}=\sum_{\mathbf{k}} \Psi^{\dag}_{\mathbf{k}}\left(
\begin{array}{cc}
 h(\mathbf{k}) & 0 \\
 0 & h^*(\mathbf{-k})\\
\end{array}
\right)\Psi_{\mathbf{k}} \label{Hk} \\
&h(\mathbf{k})=\varepsilon_\mathbf{k}I_2+d_{\alpha}(\mathbf{k})\cdot s^{\alpha}
\end{split}
\end{align}
 where $\Psi_{\mathbf{k}}=\{ C_{E,\mathbf{k},\uparrow}, C_{H,\mathbf{k},\uparrow},C_{E,\mathbf{k},\downarrow}, C_{H,\mathbf{k},\downarrow} \}^T$, $s^{\alpha}$'s are Pauli matrices,
\begin{align}
C_{\tau,\mathbf{k},\sigma}=\frac{1}{\sqrt{N}} \sum_{i}\exp(i\mathbf{k} \cdot \mathbf{R}_{i}) C_{i,\tau,\sigma}
\end{align}
where N is the total number of sites and
\begin{align}
\begin{split}
&\varepsilon_{\mathbf{k}}=C-\frac{2D}{a^2}(2-\cos(k_x)-\cos(k_y)) \\
&d_{\alpha}(\mathbf{k})=[\frac{A}{a}\sin(k_x),-\frac{A}{a}\sin(k_y),M(\mathbf{k})]\\
&M(\mathbf{k})=M-\frac{2B}{a^2}(2-\cos(k_x)-\cos(k_y)) \\
D&=(t_E+t_H)/2, B=(t_E-t_H)/2, A=2t_{EH}\\
M&=\frac{\varepsilon_E-\varepsilon_H}{2}-2(t_E-t_H)\\
C&=\frac{\varepsilon_E+\varepsilon_H}{2}-2(t_E+t_H)
\end{split}
\end{align}
 where a is the lattice constant.  Expanding the Hamiltonian (\ref{Hk}) around the $\Gamma$ point $\bf{k}=0$, we see that  $H$ reduces to the continuum BHZ Hamiltonian describing topological insulators\cite{Bernevig15122006}, with $-(D+(-) B)k^2$ being the kinetic term of the electron (E) and hole (H) orbital and M being the Dirac mass. The topological region is characterized by $M\cdot B >0$. $A$ describes the hybridization between the H and E orbital and $C$ is a overall constant energy term which will be absorbed in the chemical potential in the following.

We shall treat the interaction term in a mean-field theory where
\begin{align}
\begin{split}
n_{i,\tau,\sigma}n_{i\tau',\sigma'}&\approx \langle n_{i,\tau,\sigma} \rangle n_{i,\tau',\sigma'} + \langle n_{i,\tau',\sigma'} \rangle n_{i,\tau,\sigma}\\
 &- \langle n_{i,\tau,\sigma} \rangle  \langle n_{i,\tau',\sigma'} \rangle
 \end{split} \nonumber
\end{align}
and the mean field Hamiltonian becomes,
\begin{align}
H_{MF}=H_{BHZ}+\sum_{i,\sigma,\tau}(U_{\tau} \langle n_{i,\tau,-\sigma} \rangle +U_{EH} \langle n_{i,\bar{\tau}} \rangle) n_{i,\tau,\sigma}
\end{align}
where $\bar{E}(\bar{H})=H(E)$ $n_{i,\tau}=\sum_{\sigma}n_{i,\tau,\sigma}$ and $\langle...\rangle$ denotes (ground state) expectation value.

\begin{figure}[tbh!]
\centering
\mbox{
\subfigure[\label{schematicband1}]{\includegraphics[width=0.25\textwidth]{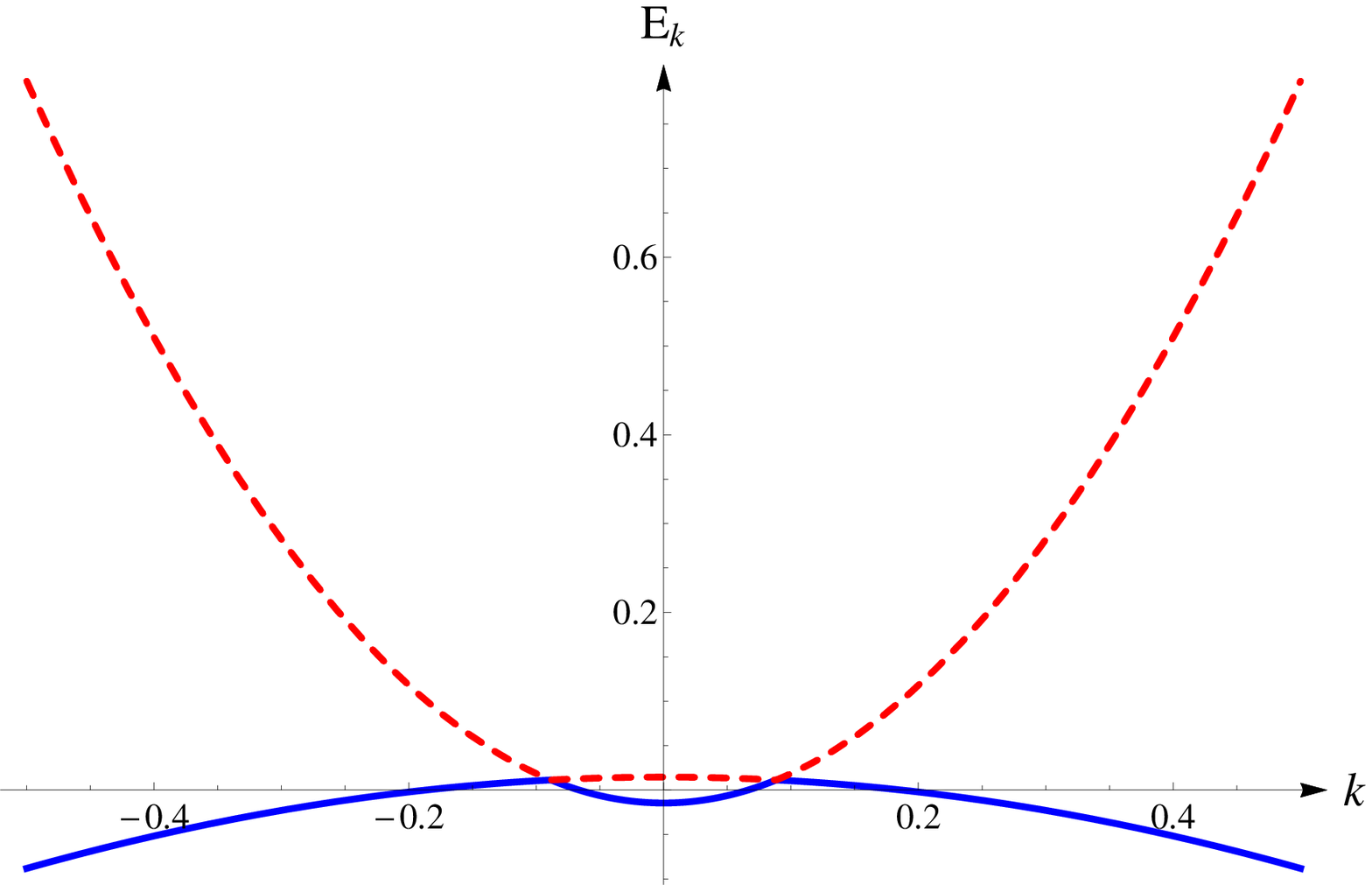}}
 \quad
\subfigure[\label{schematicband2}]{\includegraphics[width=0.25\textwidth]{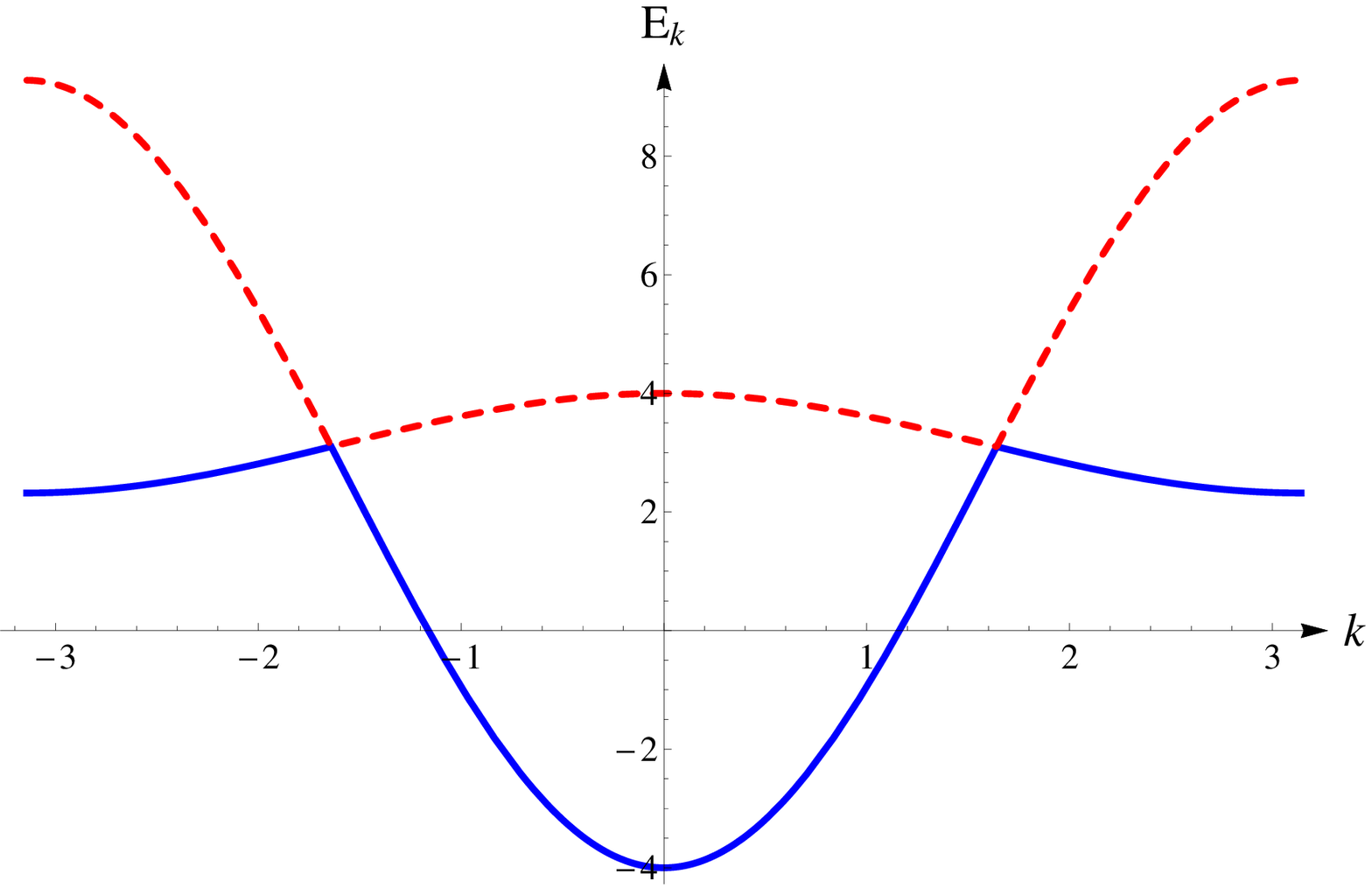}}
}
\caption{Schematic band structure illustrating the filling of the E and H bands. The blue-solid/red-dash color denotes the part of bands which are occupied/empty. (a) The case for small band inversion corresponding to HgTe/CdTe quantum well. (b) The situation with large band inversion. }.
\label{schematicband}
\end{figure}

 To understand the physics behind the mean-field theory, we first consider the case when the hybridization between the E and H orbits vanishes.  In this case the two bands overlap because of band inversion (see Fig.(\ref{schematicband})) and a small part of the E-band is occupied whereas the H-band is almost filled (see Fig.(\ref{schematicband1})). In this case the E and H bands are described separately by single-band Hubbard models which are almost empty/filled. Mean-field studies for single band Hubbard model on square lattice has been carried out long time ago\cite{PhysRevB.31.4403} and it was found that the ground state is {\em anti-ferromagnetic} at and close to half filling and becomes {\em ferromagnetic} away from half filling when the interaction strength $U$ is large than certain critical value.  In HgTe/CdTe quantum well the E and H bands are nearly empty or fully filled, correspond to the case in Fig.(\ref{schematicband1}), suggesting that we should look for ferromagnetic phases in our mean-field theory. Anti-ferromagnetic phase is expected only if the band inversion is so large that the two bands are both nearly half filled (case shown in Fig.(\ref{schematicband2})). We've searched for both {\em ferromagnetic} and {\em anti-ferromagnetic} phases numerically in our study and find that the {\em antiferromagnetic} phase has higher energy for band parameters appropriate for HgTe/CdTe quantum well. In the following we shall focus on the {\em paramagnetic} and {\em ferromagnetic} phases.

The mean field Hamiltonian which captures both {\em paramagnetic} and {\em ferromagnetic} phases is in $k$-space,
\begin{align}
H_{MF}&=H_{BHZ}+\sum_{\tau,\sigma}h_{\tau,\sigma}n_{\tau,\mathbf{k},\sigma}\\  \nonumber
h_{\tau,\sigma}&= \frac{U_{\tau}}{2}(n_{\tau}-\sigma m_{\tau})+U_{EH}n_{\bar{\tau}}
\end{align}
where $n_{\tau}$ and $m_{\tau}$ are the occupation number and the magnetization of the $\tau$ orbital, respectively with
\begin{align}
\begin{split}
&n(m)_\tau={1\over V}\sum_\mathbf{k}<n_{\tau,\mathbf{k},\uparrow}>+(-)<n_{\tau,\mathbf{k},\downarrow}>\\
&=\frac{1}{V}\sum_{\mathbf{k}}\left(\Theta(\mu-E_{\mathbf{k},\uparrow,+})|\alpha_{\tau,\mathbf{k},\uparrow}|^2+\Theta(\mu-E_{\mathbf{k},\uparrow,-})|\beta_{\tau,\mathbf{k},\uparrow}|^2\right) \\
&+(-)\left(\Theta(\mu-E_{\mathbf{k},\downarrow,+})|\alpha_{\tau,\mathbf{k},\downarrow}|^2+\Theta(\mu-E_{\mathbf{k},\downarrow,-})|\beta_{\tau,\mathbf{k},\downarrow}|^2 \right)
\\
&|\alpha_{\tau,\mathbf{k},\sigma}|^2=\frac{1}{2}(1+\frac{\text{Sign}_{\tau} M_{\mathbf{k}}}{\sqrt{(M_{\mathbf{k}}+\Sigma_{2\sigma})^2+|A_{\mathbf{k}}|^2}})\\
&|\beta_{\tau,\mathbf{k},\sigma}|^2=\frac{1}{2}(1-\frac{\text{Sign}_{\tau}M_{\mathbf{k}}}{\sqrt{(M_{\mathbf{k}}+\Sigma_{2\sigma})^2+|A_{\mathbf{k}}|^2}})
\end{split}
 \label{nh}
\end{align}
where $n_{\tau,\mathbf{k},\sigma}=C_{\tau,\mathbf{k},\sigma}^\dag C_{\tau,\mathbf{k},\sigma}$  and $\text{Sign}_{E(H)}=1(-1)$. $\Theta(x)$ is the step function.  $E_{\mathbf{k},\sigma,\pm}$ is the eigen-energies of $H_{MF}$ given by,
\begin{align}
&E_{\mathbf{k},\sigma,\pm}=\varepsilon_\mathbf{k}+\Sigma_{1\sigma}\pm \sqrt{(M_\mathbf{k}+\Sigma_{2\sigma})^2+|A_\mathbf{k}|^2}
\end{align}
where,
\begin{align}
\Sigma_{1\sigma}&=\frac{U_E(n_E-\sigma m_E)+U_H(n_H-\sigma m_H)+4U_{EH}}{4},\\  \nonumber
\Sigma_{2\sigma}&=\frac{U_{E}(n_E-\sigma m_E)-U_H(n_H-\sigma m_H)+2U_{EH}(n_H-n_E)}{4},\\ \nonumber
A_{\mathbf{k}}&=\frac{A}{a}(\sin(k_x)+i \sin(k_y)).
\end{align}
 The ground state energy density is,
\begin{align}
{E_g\over V}&={1\over V}\sum_{\mathbf{k}}\sum_{i=\pi,\sigma} \Theta(\mu-E_{\mathbf{k},i})E_{\mathbf{k},i}\\  \nonumber
&-\frac{U_E}{4}(n_E^2-m_H^2)-\frac{U_H}{4}(n_H^2-m_H^2)-U_{EH}n_E n_H.
\end{align}

The mean field parameters $n(m)_{\tau}$ are determined self-consistently {\em via} the mean-field equations.

\section{Results and phase diagram}
We consider the half filled case for the BHZ model where the chemical potential is in the gap and the system is a topological insulator. We employ the parameters appropriate for the 7.5nm HgTe/CdTe quantum well with $M=-0.0146 eV,~B=-1.87 eV\cdot nm^2, ~ A=0.55 eV \cdot nm, ~ D=-1.45 eV \cdot nm^2$ and  $C=0$. \cite{ReentrantTP}. The lattice constant $a$ is chosen to be 1 nm.

\begin{figure}[tbh!]
\centering
\mbox{
\subfigure[\label{orderparameter}]{\includegraphics[width=0.22\textwidth]{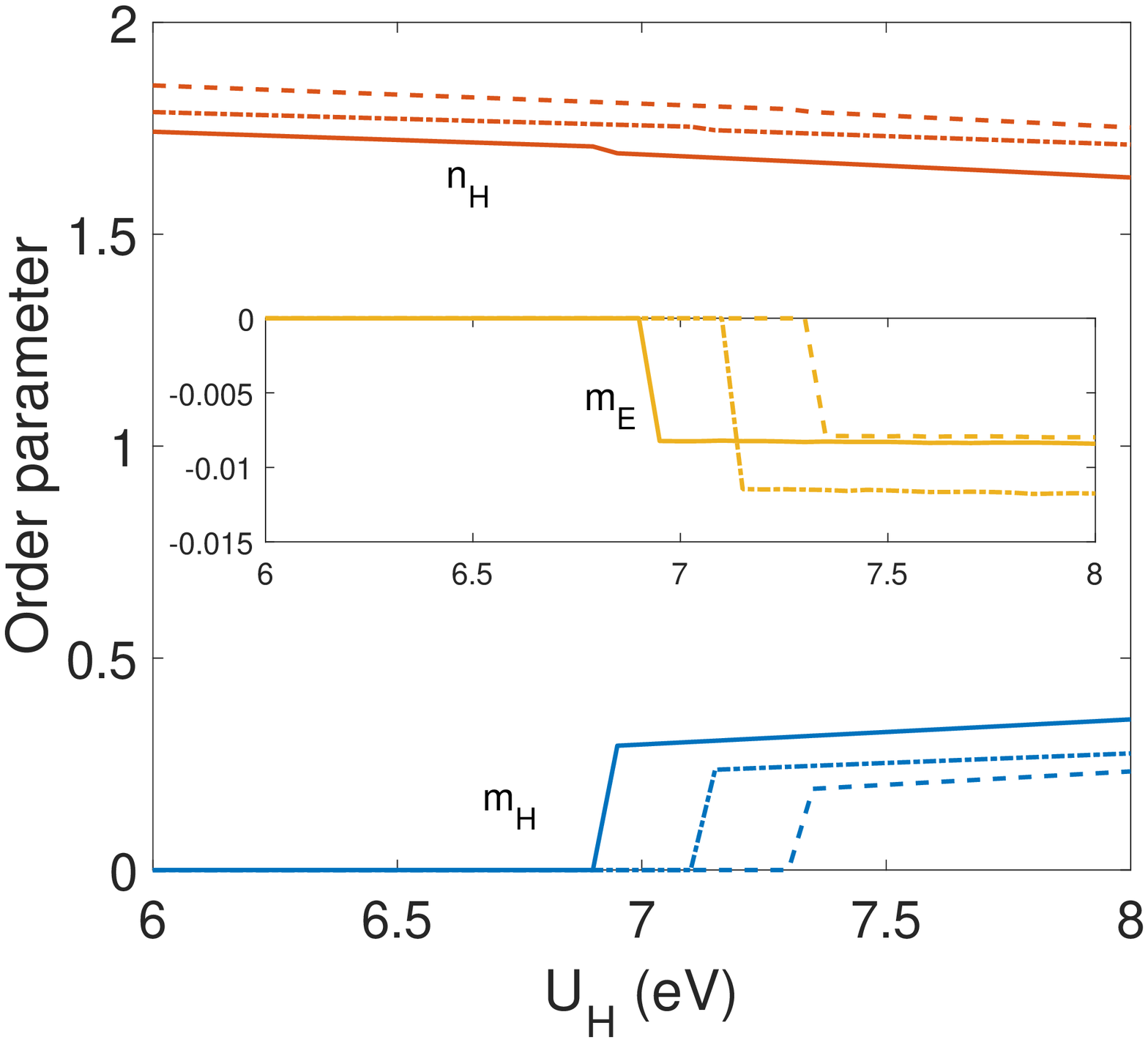}}
 \quad
\subfigure[\label{energyd}]{\includegraphics[width=0.22\textwidth]{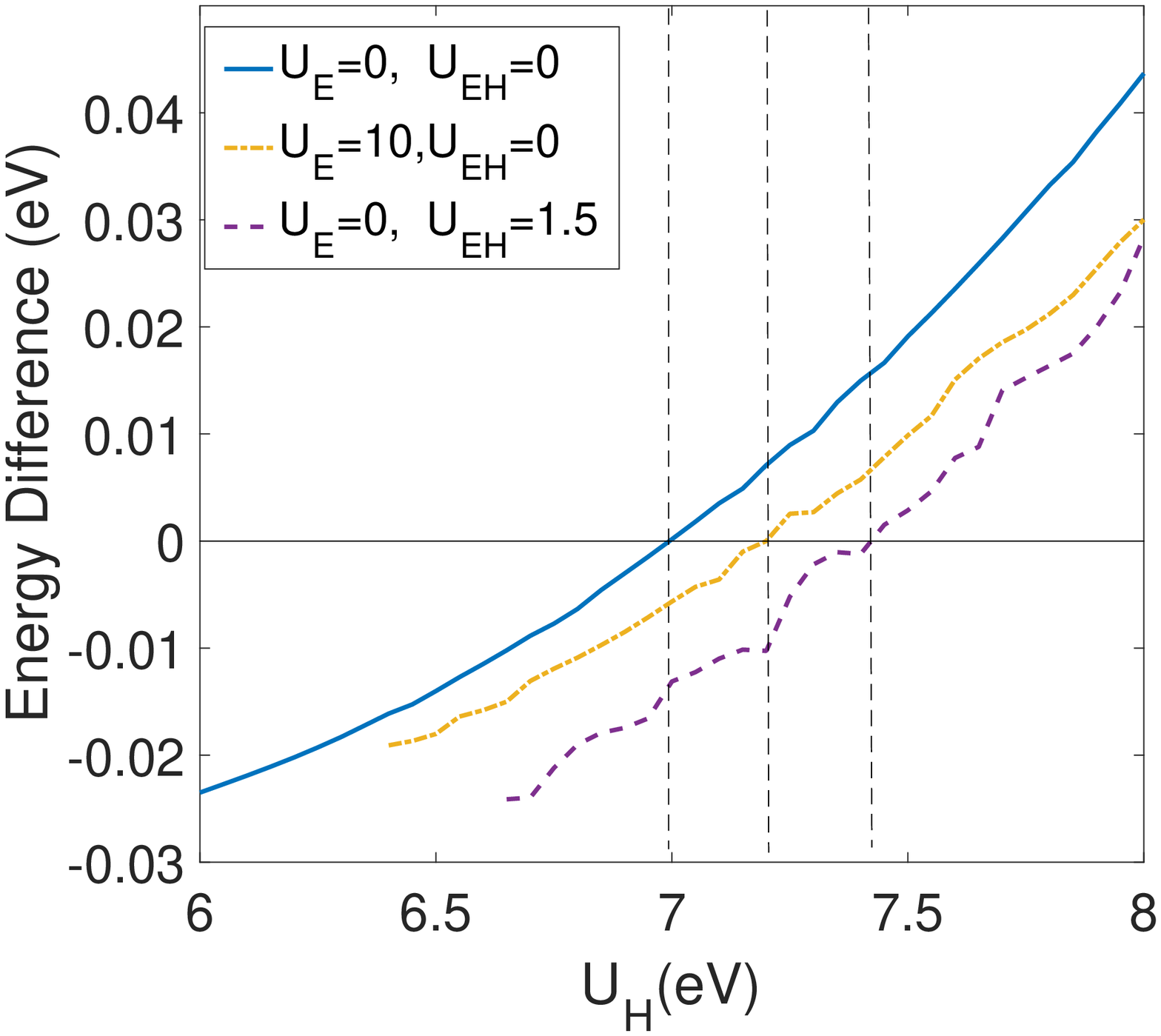}}
}
\caption{(a). Order parameters $n_H,m_H,m_E$ verse interaction strength $U_H$. Solid lines  are for $U_E=0,U_{EH}=0$. Dash-dot lines  are for $U_E=10 eV,U_{EH}=0$. Dash lines are for $U_E=0 eV,U_{EH}=1.5 ~eV$. (b) Energy difference between nonmagnetic phase (NM) and { \em ferromagnetic}  phase (FM) $E_{NM}-E_{FM}$. Different styles of line represent same parameter in (a)}.
\end{figure}

In Fig.(\ref{orderparameter}) we show the self-consistent determined mean field parameters, $n_H$, $m_E$ and $m_H$ as a function of interaction strength $U_H$ of the H orbit with (i) $U_{E}=U_{EH}=0$, (ii) $U_E=10 eV, U_{EH}=0$ and (iii) $U_{E}=0,U_{EH}=1.5 eV$ respectively. We first consider the case with only $U_H\neq0$ which is similar to the single band Hubbard model. We note that an important difference between the single band Hubbard model and the BHZ model is that in our case, the relative position of the two bands depends on interaction.  When $U_H$ increases, the on-site energy of H orbital is shifted upward while the E orbital energy remains stationary leading to increasing population in E band. Changing other interactions have similar effects. Thus we are actually moving along a curve in the density-interaction phase diagram of the one-band Hubbard model when interaction changes.  At small $U_{H}$, only one solution with $m_{H}=0$ is found. The system remains a TI. As interaction strength increase, two self-consistent solutions appear. The ground state is the one with lower energy.  The energy difference between the {\em ferromagnetic} and{\em paramagnetic } phases are shown in Fig.(\ref{energyd}). We find a first order phase transition between {\em paramagnetic} and {\em ferromagnetic} phases indicated by the sign change of the energy difference between the two phases. We note that despite $U_E=0$ in this case, $m_E$ has non-vanishing negative value as long as $m_H\neq 0$ because of nonzero band hybridization $A\neq0$.

\begin{figure}[b]
\centering
\mbox{
\subfigure[\label{B0}]{\includegraphics[width=0.2\textwidth]{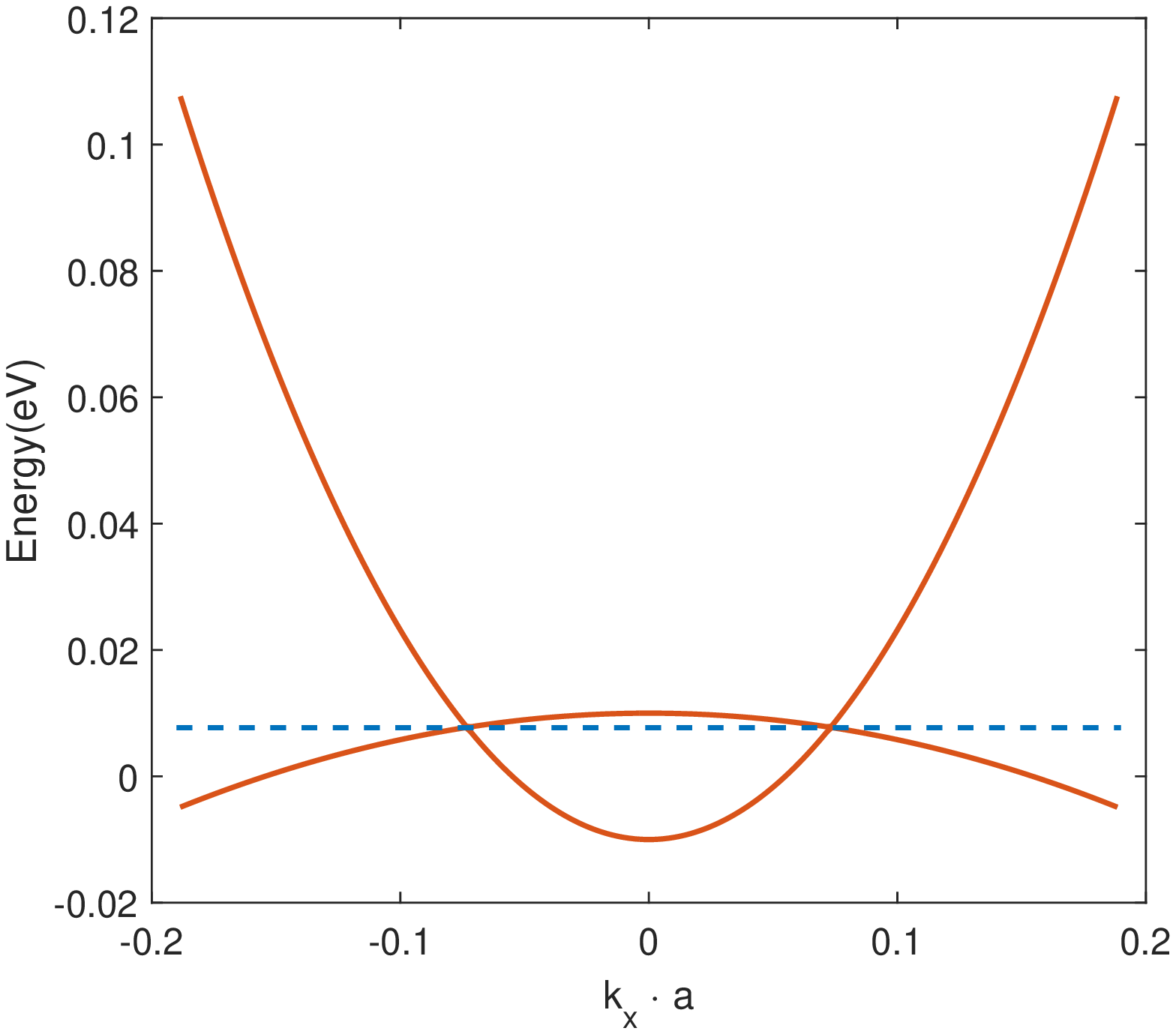}}
 \quad
\subfigure[\label{Bc}]{\includegraphics[width=0.2\textwidth]{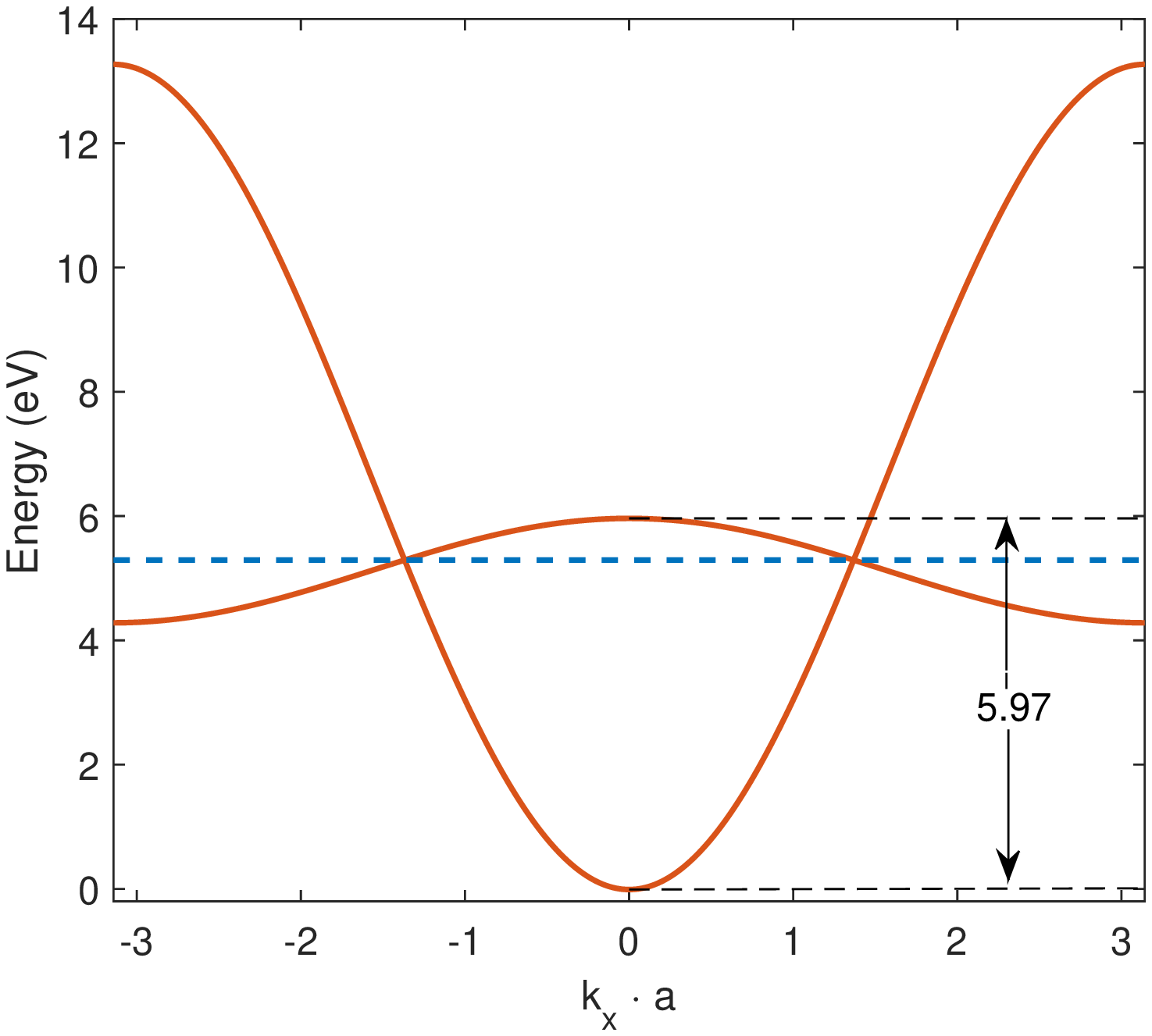}}
 \quad
}
\mbox{
\subfigure[\label{BE10}]{\includegraphics[width=0.2\textwidth]{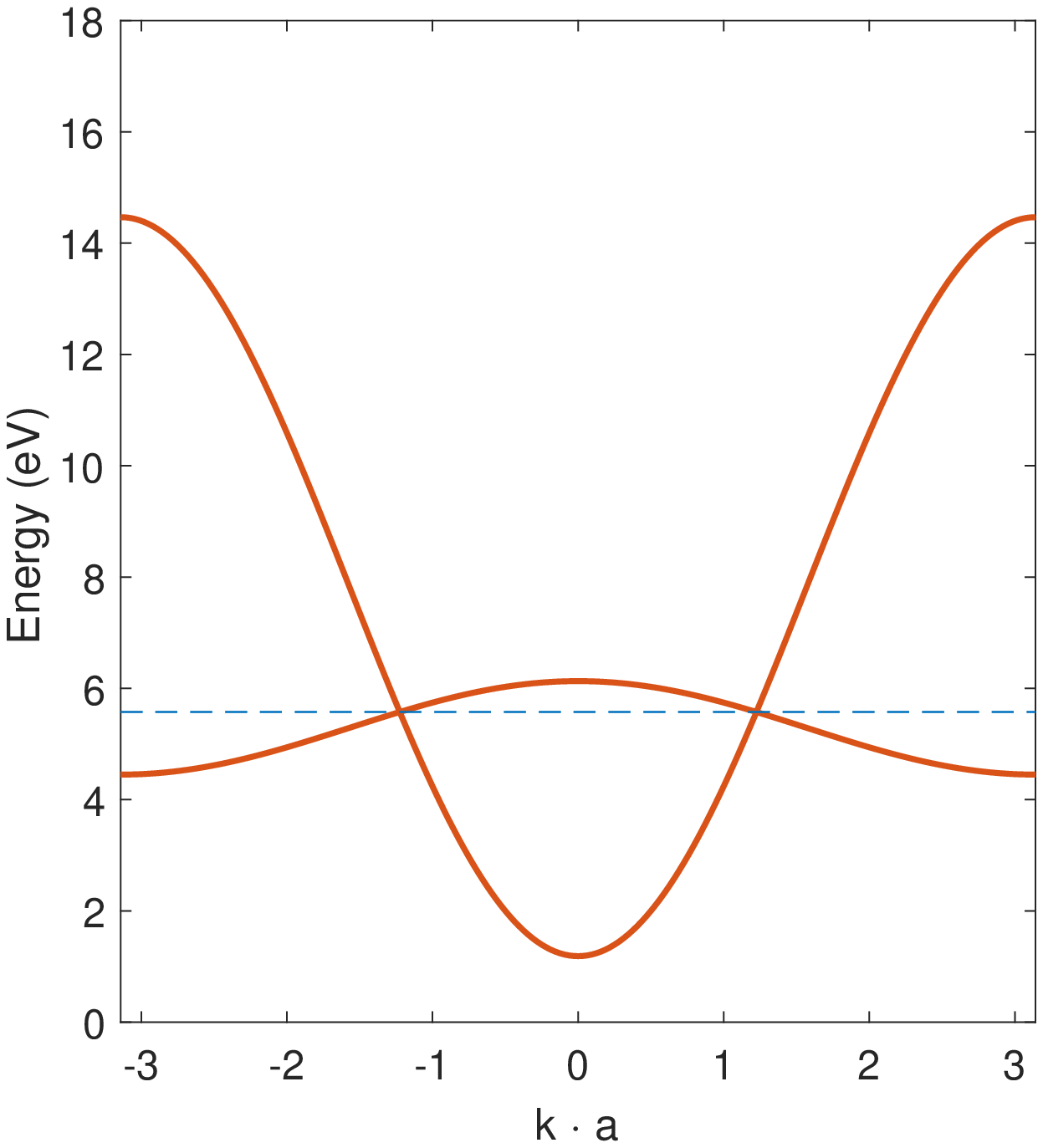}}
 \quad
\subfigure[\label{BEH}]{\includegraphics[width=0.2\textwidth]{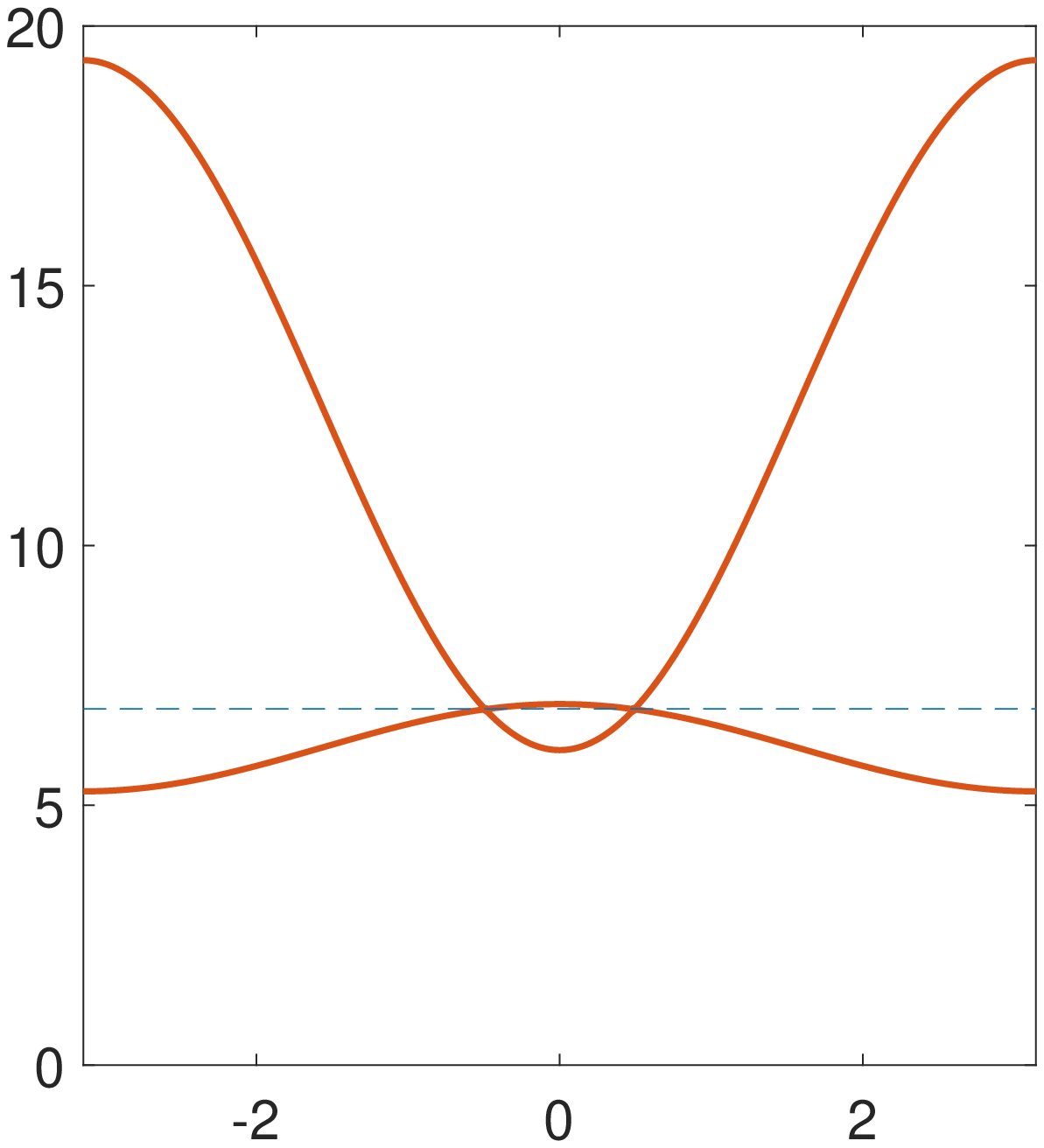}}
 \quad
}
\caption{Band structure assuming $A=0$ (along $k_x \in [-\pi, \pi] $,$k_y=0$) evolution with interaction strength. (a) $U_H=U_{E}=U_{EH}$. The blue-dash line shows the chemical potential. (b) $U_H=6.95 eV, ~ U_{E}=E_{EH}=0$. (c) $U_H=6.95 eV, ~ U_{E}=10 eV,~ E_{EH}=0$ . (d) $U_H=6.95 eV, ~ U_{E}=10 eV,~ E_{EH}=3 eV$}
\label{BSpicture}
\end{figure}
Including $U_E$ and $U_{EH}$ have similar effect as $U_H$. $U_E$ increases the energy of E orbital (see Fig.(\ref{BSpicture}). However, as discussed above, the occupation number of E orbital is much smaller than H, and the effect of $U_E$ is much smaller compared to $U_H$ because of smallness of $n_E$.  Therefore $U_E$ has almost no effect on the phase transition.  $U_{EH}$ raises the energies of the two orbital simultaneously but with different values depending on the occupation numbers of the two bands. The shift in energy of E(H) orbital is proportional to $n_H(n_E)$. Again, since $n_H>> n_E$, the energy of E orbital is shifted faster than H orbital leading to decreasing/increasing occupation number in E/H orbital for $U_{EH}>0$.

\begin{figure}[tbh!]
\centering
\mbox{
\subfigure[$U_{E}=0$]{\includegraphics[width=0.23\textwidth]{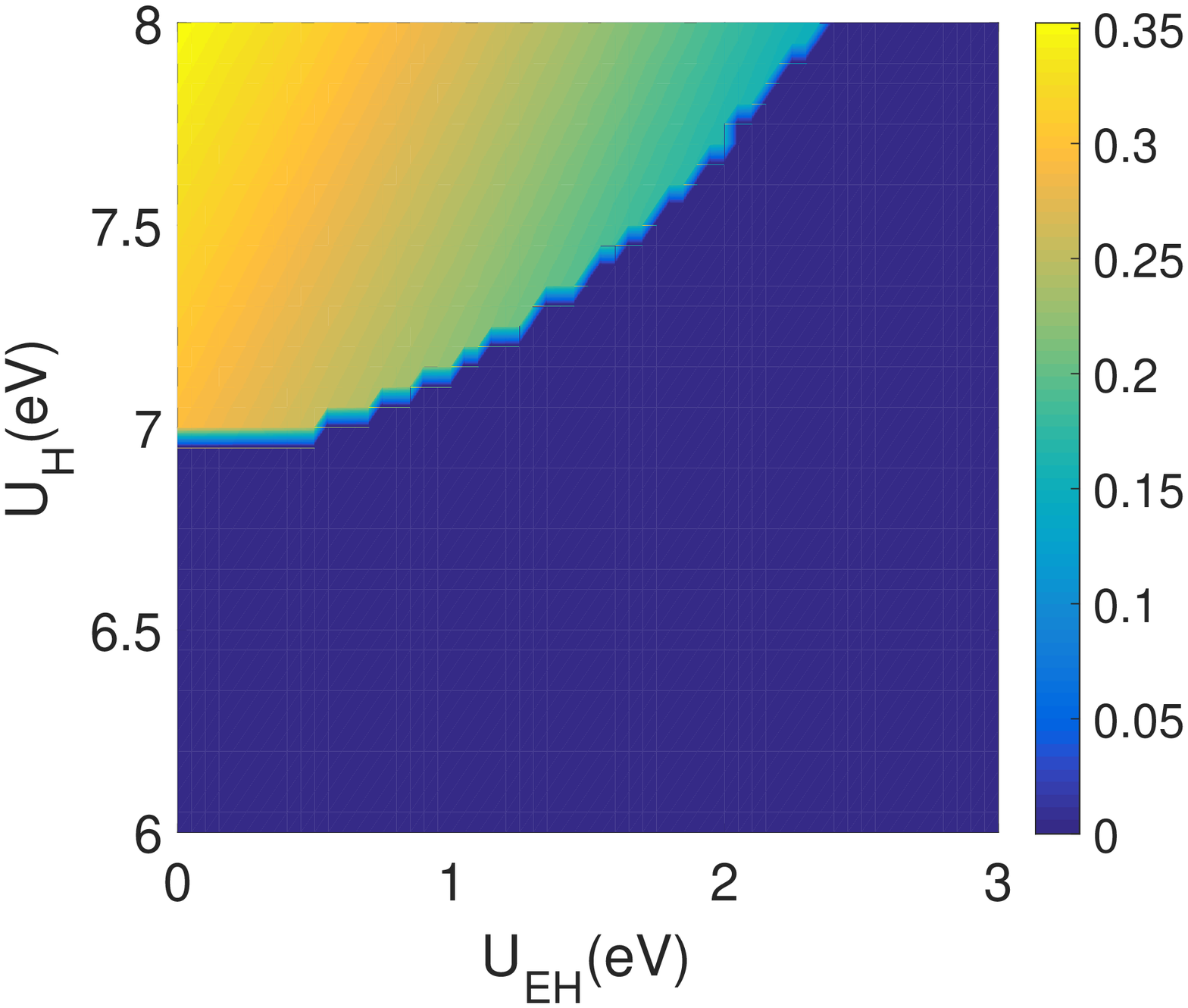}}
 \quad
\subfigure[$U_{E}=10 ~eV $]{\includegraphics[width=0.23\textwidth]{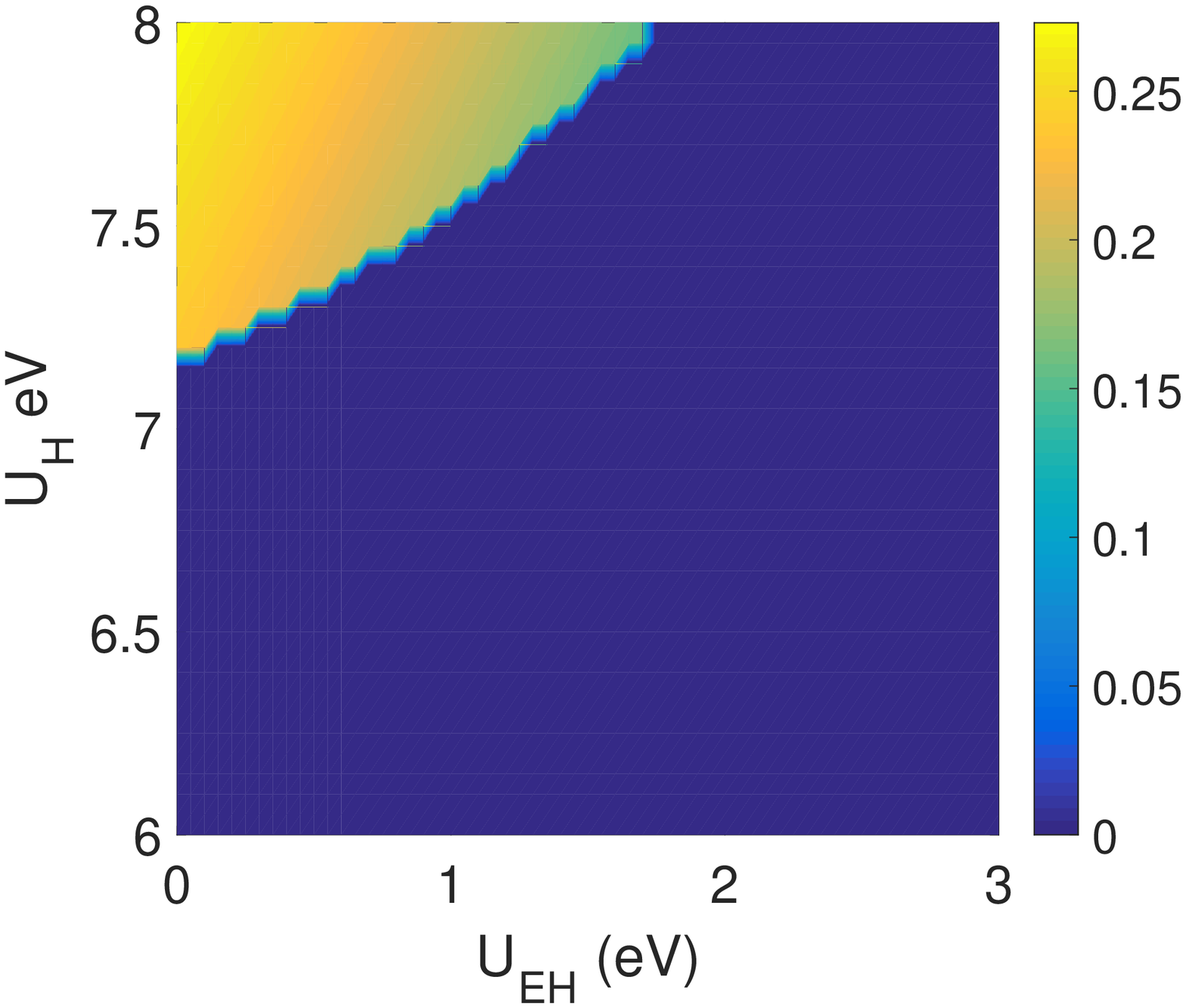}}
 \quad
}
\caption{Phase diagram in $U_H-U_{EH}$ space with given $U_E$. The color represent the magnitude of magnetization.}
\label{phasediagram}
\end{figure}

 The dependence of the {\em paramagnetic-ferromagnetic} phase boundary on the interactions are summarized in the phase diagram in Fig.(\ref{phasediagram}). we see that the ferromagnetic region shrink when $U_E, U_{EH}>0$ since these interactions effectively increase/decrease the occupation of H/E orbital. We note that besides the {\em ferromagnetic} phase, we have also searched for the solution of {\em anti-ferromagnetic} phase. We find that the anti-ferromagnetic phase exists only if $U_{H}>>U_{E},U_{EH}$ such that the energy of H orbital is lifted to be close to the 1/2 filling case shown in Fig.(\ref{schematicband2}).

\begin{figure}[tbh!]
\centering
\mbox{
\subfigure[\label{B1}]{\includegraphics[width=0.12\textwidth]{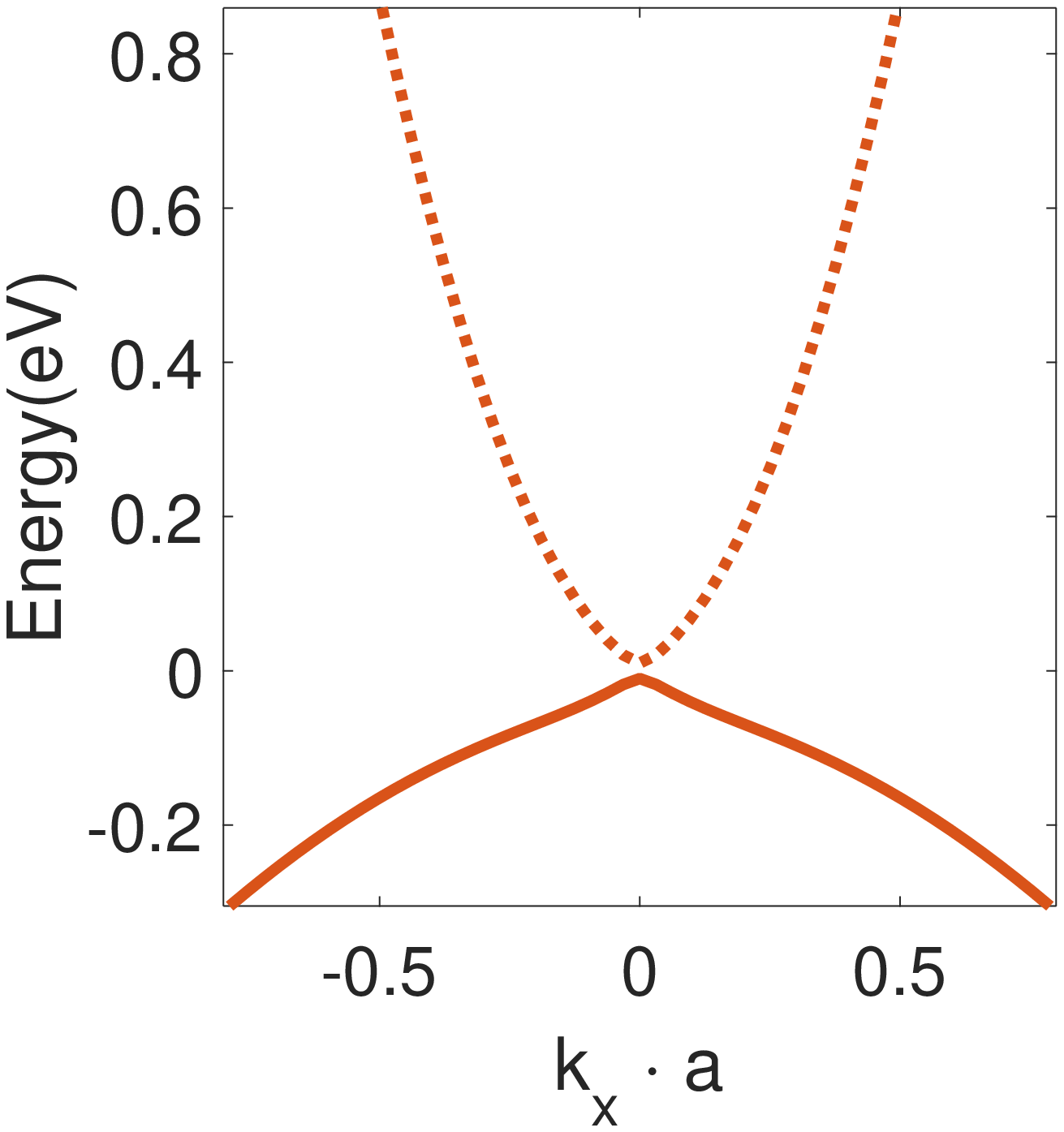}}
 \quad
\subfigure[\label{B2}]{\includegraphics[width=0.12\textwidth]{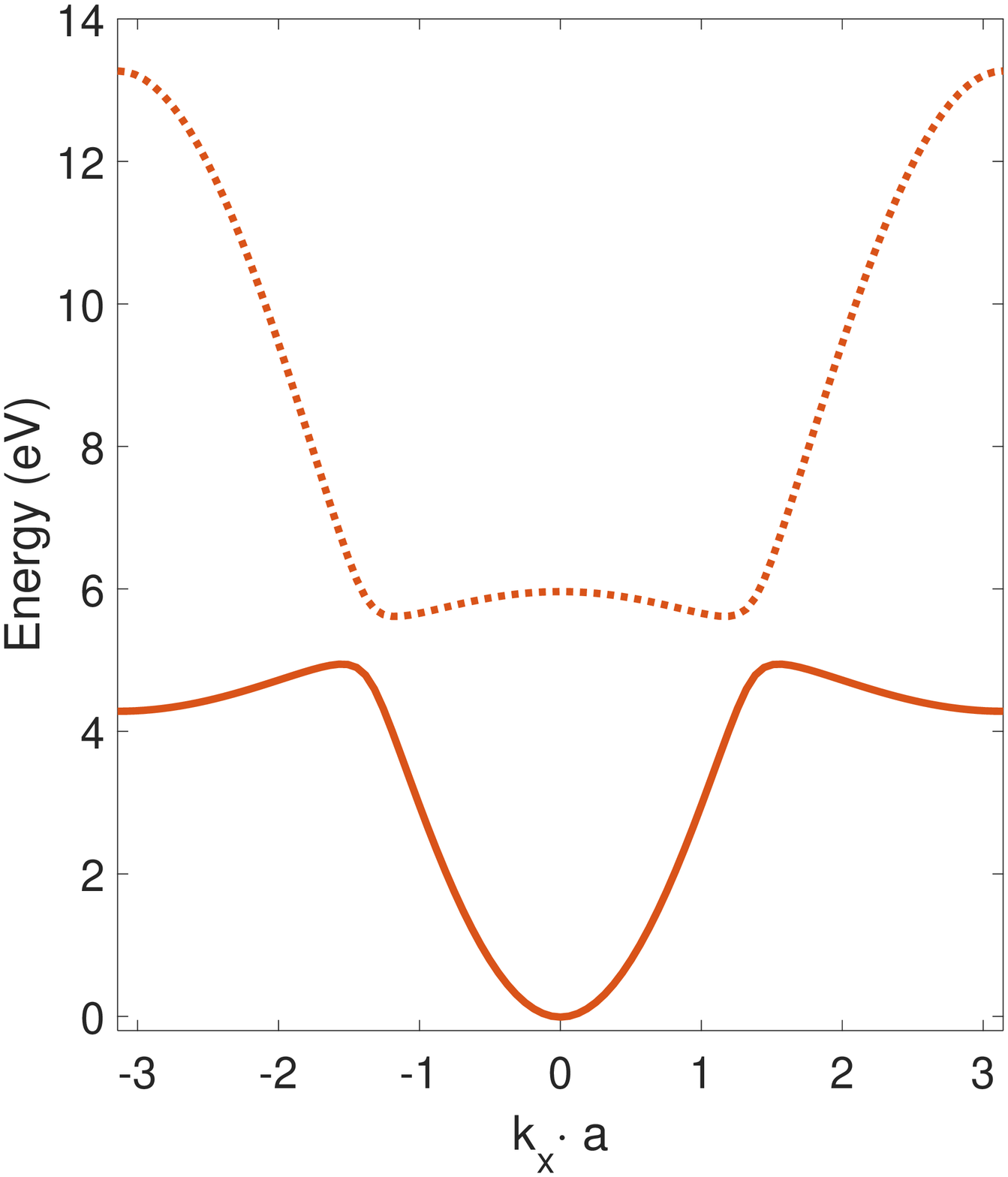}}
 \quad
\subfigure[\label{B3}]{\includegraphics[width=0.12\textwidth]{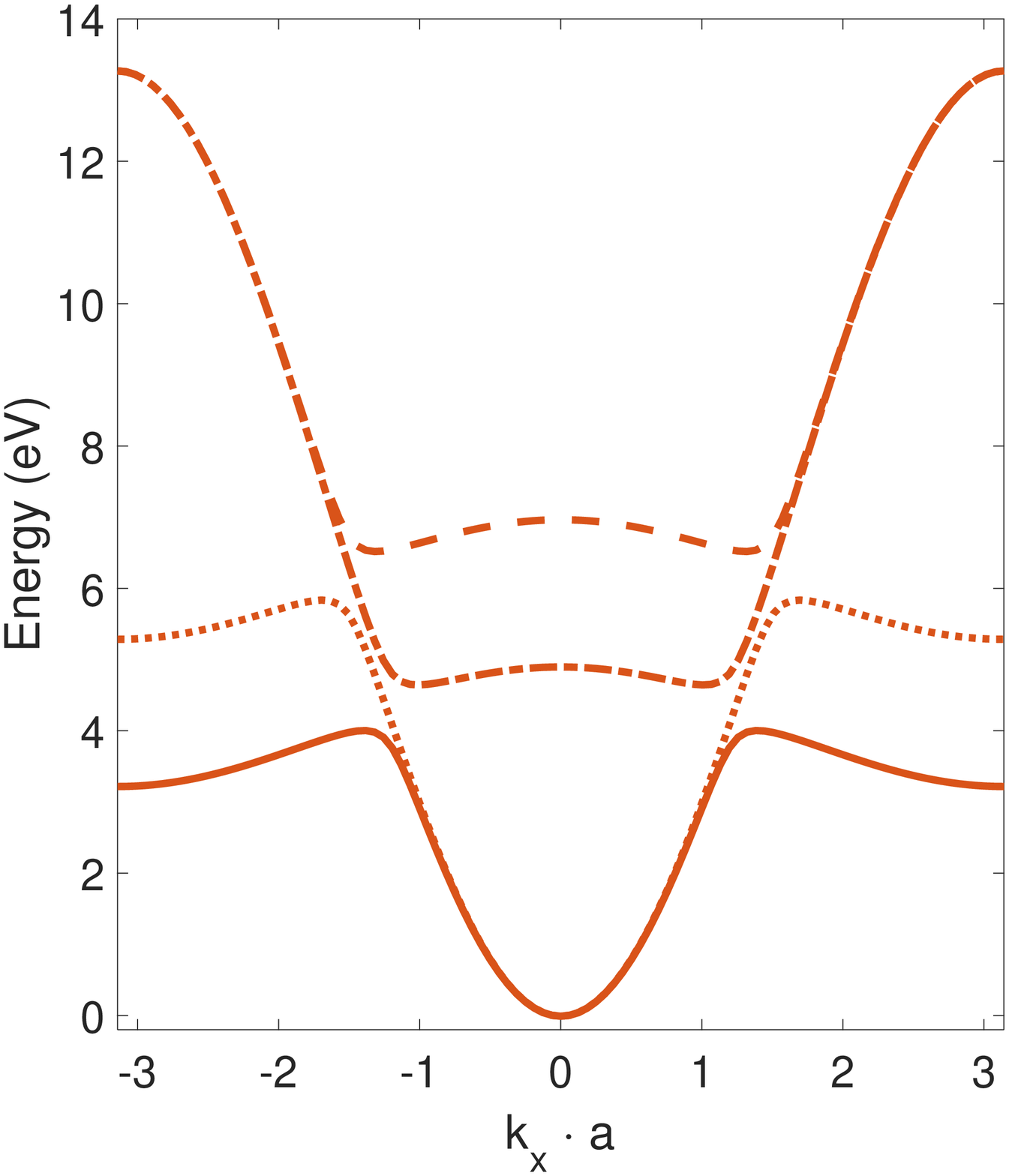}}
}
\caption{Band structure (along $k_x \in [-\pi, \pi] $,$k_y=0$) evolution with interaction strength $U_H$. }
\label{BS}
\end{figure}

The evolution of the band structure in the two phases is shown in Fig.(\ref{BS}). The band structure without interaction is shown in Fig.(\ref{B1}). In this case the bands are doubly degenerate with band extremum at $\Gamma$ point. The band structure near the phase transition point ($U_{H}=6.95 eV, ~U_{E}=U_{H}=0$) is shown in Fig.\ref{B2}. We see that the band extremum are shifted to finite $k$-points due to the upward shifting of H-orbital compared with the E-orbital. The band degeneracy is lifted by the magnetization and the system becomes gapless in the {\em ferromagnetic} phase (Fig.\ref{B3}).
We caution that the E and H orbital are linear combination of different atomic orbital (s and p orbits) in realistic material.  Although a state with uniform magnetization is obtained in the lattice model, the magnetization distribution on the two atomic orbital are in general different in real materials.

\section{Summary and discussions}
Summarizing, we studied in this paper the effects of Hubbard-type on-site interaction on the BHZ model. Within a mean-field theory, we find a ferromagnetic state when $U_H$ is large enough and the band inversion is small. The system becomes {\em gapless} (metallic) in the ferromagnetic phase. We have computed the Chern number (C) for the energy bands in the ferromagnetic phase and find that they have values $C=\pm 1$, i.e., the ferromagnetic transition is {\em not} a topological phase transition.

Our result is qualitatively consistent with the phase diagram inferred from Ref.[28] for the single-band Hubbard model. The magnitude of $U_H$ is roughly 7 eV close to the critical point and is about a quarter of the band width, which is not unrealistic. It should be emphasized that mean field approximation has overestimated the tendency of {\em ferromagnetic} ordering. The region of {\em ferromagnetic} phase shrinks when more accurate calculations are performed on the one-band Hubbard model\cite{bookmag,JK} and similar result is expected here. We note that more exotic phases are proposed to appear in the one band Hubbard model when the system is close to half-filling\cite{PhysRevLett.95.237001,RevModPhys.66.763,PhysRevLett.85.5162,PhysRevLett.62.591,0295-5075-4-5-016,Yoshioka1989,PhysRevB.70.125111,PhysRevLett.97.036401,PhysRevLett.64.1445,PhysRevB.74.024508,PhysRevB.42.10641,PhysRevB.75.075110,PhysRevB.62.R9283,doi:10.1143/JPSJ.66.294,PhysRevB.74.054513}
and similar situation is expected here when the H-orbital is lifted up further such that both E- and H- orbital are close to half-filling. The parameters we use in this paper are base on 7.5nm HgTe/CdTe quantum well.

Our result may be generalized to other similar 2DTI material; for example, the InAs/GaSb quantum well. Due to the inversion symmetry breaking, InAs/GaSb quantum well has considerable Rashba spin orbit coupling and is described by a modified BHZ model on square lattice with extra (small) bulk inversion asymmetry (BIA) and structural inversion asymmetry (SIA) term described by the tight-binding Hamiltonian $H_{TB}=H_{BHZ}+H_{BIA}+H_{SIA}$, where
\begin{align}
H_{BIA}=&\sum_{\mathbf{k},\tau}\Delta_{E}(\mathbf{k}) C_{E,\mathbf{k},\uparrow}^{\dag}C_{E,\mathbf{k},\downarrow}+\Delta_{H}(\mathbf{k})C_{H,\mathbf{k},\uparrow}^{\dag}C_{H,\mathbf{k},\downarrow}\nonumber \\
&+\Delta_{0}(C_{E,\mathbf{k},\uparrow}^{\dag}C_{H,\mathbf{k},\downarrow}-C_{H,\mathbf{k},\uparrow}^{\dag}C_{E,\mathbf{k},\downarrow})+c.c
\\ \nonumber
H_{SIA}=&\sum_{\mathbf{k}}i\xi_{E}(\mathbf{k})C_{E,\mathbf{k},\uparrow}^{\dag}C_{E,\mathbf{k},\downarrow}+c.c
\end{align}
where $\Delta_{E(H)}(\mathbf{k})\to \Delta_{E(H)} k_{+(-)}, \xi_{E}(\mathbf{k})\to \xi_{E}k_{-}$ in the $\mathbf{k}
\to 0$ limit. $\xi_{E}$ is the electron Rashba term where $k_{\pm}=k_x\pm i k_y$.
 This leads to spin-flip term in the Hamiltonian $H_{TB}$. However, the interaction term is of the same form as the $H_U$ and the Hamiltonian has the same form as $H_T$ after diagonalizing $H_{TB}$ except that the spins have different quantization axis at each k point. In particular, the Hubbard interaction has similar effect on the model as the original Hubbard-BHZ model and the effect of orbital filling will be similar.

As discussed above, we see that when the H-orbital is lifted up and the system becomes a large gap topological insulator, stronger effects of electron interaction (and more exotic phases) are expected when the orbital are close to half-filling. The search for large gap topological insulator has been a hot topic after the discovery of topological insulator  since larger gap implies that the topological effects can be measured at higher temperatures. This is important for the application of TI based electronic devices. Our calculation suggests that large gap TI is also helpful to realize exotic phases in TI.  We note that large gap 2DTI has been predicted to exist in transition metal dichalcogenides\cite{Qian1344} described by the  $\mathbf{k}\cdot\mathbf{p}$  Hamiltonian,
\begin{align}
H_{TMD}=\left(
\begin{array}{cccc}
E_{p}(k_x,k_y) & 0 & -i v_{1}\hbar k_{x} & v_2 \hbar k_{y}\\
0 & E_{p}(k_x,k_y) & v_{2}\hbar k_{y} & -v_1 \hbar k_x\\
iv_1 \hbar k_x &v_2 \hbar k_y & E_{d}(k_x,k_y) & 0\\
v_2 \hbar k_y & i v_1 \hbar k_x & 0 & E_{d}(k_x,k_y)
\end{array}
\right)
\end{align}
 in the basis of p and d orbital where $E_{p}=-\delta+\frac{\hbar^2 k_{x}^2}{2m_{x}^{p}}+\frac{\hbar^2 k_y^{2}}{2m_{y}^{p}},E_{d}=\delta+\frac{\hbar^2 k_{x}^2}{2m_{x}^{d}}+\frac{\hbar^2 k_y^{2}}{2m_{y}^{d}}$. $\delta$ is the inverted gap with value ranges from 0.284 eV to 0.978 eV (calculated in $GW$ approximation) \cite{Qian1344}. The inverted gaps are  20 to 100 times of HgTe/CdTe and InAs/GaSb quantum wells leading to closer to half band fillings for both p and d orbital. There has been some experimental evidence indicating that the 1T' structure of TMD material $\rm WTe_2$  is a 2D topological insultor\cite{fei2016topological,zheng2016quantum}. Our study suggests that this material may be a good candidate for strong correlation effect.

\section{acknowledgement}
This work is supported by Hong Kong RGC through grant HKUST3/CRF/13G.
\bibliographystyle{h-physrev2}
\bibliography{ref}
\end{document}